\documentclass[preprint2, times]{aastex631}

\newcommand{\roig}{\emph{Roig1 Prades Sky}}
\newcommand{\mast}{\,\rm{mag}\,\rm{arcsec}^{-2}}

\begin{document}

\title{Discovery of a large and faint nebula at the Triangulum galaxy}

\author{Aleix Roig}
\affiliation{Parc Astron\`omic de les Muntanyes de Prades, Muralla 5, 43364 Prades, Tarragona, Spain}
\email{info@astrocat.info}

\author[0000-0002-6220-7133]{Ra\'ul Infante-Sainz}
\affiliation{Centro de Estudios de F\'isica del Cosmos de Arag\'on (CEFCA), Plaza San Juan, 1, E-44001, Teruel, Spain}

\author[0000-0002-4383-0229]{Judith Ardèvol}
\affiliation{Institut de Ci\`encies del Cosmos (ICCUB), Universitat de Barcelona (UB), Mart\'i i Franqu\`es 1, E-08028 Barcelona, Spain}
\affiliation{Dept. de Física Qu\`antica i Astrof\'isica (FQA), Universitat de Barcelona (UB), Mart\'i i Franqu\`es 1, E-08028 Barcelona, Spain}
\affiliation{Institut d'Estudis Espacials de Catalunya (IEEC), c. Gran Capit\`a, 2-4, 08034 Barcelona, Spain}


%% Abstract environment. 
\begin{abstract}
We report the discovery of a previously uncatalogued arch-shaped filamentary nebula at the outer part of the Triangulum galaxy (M33) centred at $\alpha = 1^{h}34^{m}25^{s}$, $\delta = +30^{\circ}20'17''$ (ICRS).
This discovery stems from meticulous observations employing deep exposures of M33, using both H$\alpha$ and [OIII] narrow-band filters.
The nebula, designated as \roig, exhibits an H$\alpha$ surface brightness of $23.9\mast$.
Its sky projected location is $21\,$arcmin away from the M33 galactic centre towards the southeast direction with an extent of $(120\times440)\pm30\,$pc.
Deep spectroscopic observations are required to unveil its real nature.
\end{abstract}



\section{Introduction}
\label{sec:intro}
Amateur astronomers have more access to telescope observation time in comparison to their professional counterparts, enabling them to conduct longer observations covering wider areas of the sky.
Aleix Roig used this advantage to capture a very deep image of M33 covering an area of $2.4^{\circ}\times1.6^{\circ}$ around this galaxy.
The primary objective was to seek for [OIII] and H$\alpha$ emission regions in the outskirts of M33.

These observations took place at the Astrocat Observatory\footnote{\url{https://astrocat.info}}, an amateur facility located in the dark skies of Prades (Spain), characterised by a sky brightness of $\mu_{\rm{V}} \sim 21 \mast$. 
The exceptional conditions allowed Aleix Roig to set up this observatory back in 2014.
%It consists of two refractor telescopes (FSQ85 and FSQ106) as well as another small support refractor.
He held other professional collaborations, such as the 100\,hours image of the M101 galaxy presented at the 355 IAU Symposium\footnote{\url{https://www.iac.es/es/node/188587}}.

\section{Data}
\label{sec:data}
The images were acquired employing two small refractor telescopes, both synchronised by a MESU200 mount.
Each telescope is equipped with its independent camera and filters, configured to observe the same region of the sky.
The images were taken during two distinct periods: from September 22 to 26, and from October 5 to 11, 2023. Periods of high Moon illumination were carefully avoided.

The first telescope, a Takahashi FSQ85 (85\,mm aperture, F/5.3), captured a total of 969 images, each lasting 300 seconds (total exposure time of 80.75\,hours) using the [OIII] filter centred at 500.7\,nm with a passband of 6.5\,nm.
We used a ZWO ASI294MM 4/3" camera with a pixel size of 2.1\,arcsec/pixel and a field of view (FOV) of $2.44^{\circ}\times1.66^{\circ}$.

The second telescope, a Takahashi FSQ106 (106\,mm aperture, F/5.0), captured a total of 534 images, each lasting 300\,seconds (total exposure time of 44.5\,hours) using the H$\alpha$ filter centred at 656.3\,nm with a passband of 3\,nm.
The camera used was a ZWO ASI2600MM APS-C with a pixel size of 1.5\,arcsec/pixel and a FOV of $2.54^{\circ}\times1.74^{\circ}$.

The Takahashi FSQ106 was used to capture images with luminance (L), red (R), green (G), and blue (B) filters that were combined to obtain an LRGB-coloured image with a total exposure time of 28.5\,hours.

The data processing was mainly done with \texttt{PixInsight}\footnote{\url{https://pixinsight.com/}} to obtain the colour images.
The use of J-PLUS \citep{cenarro2019} data, and more in particular its H$\alpha$ (J0660) filter, allowed us to perform a crude photometric calibration and the estimation of surface brightness measurements by using \texttt{Gnuastro} \citep{akhlaghi2015}.
The depth of the combined H$\alpha$ image is $24.9\mast$ ($3\sigma$ in 100\,arcsec$^2$).
Due to the lack of good-quality calibrated data, we could not obtain reliable brightness measurements in the [OIII] filter.

\section{Results}
\label{sec:results}
\begin{figure*}[t]
    \includegraphics[width=0.99\textwidth]{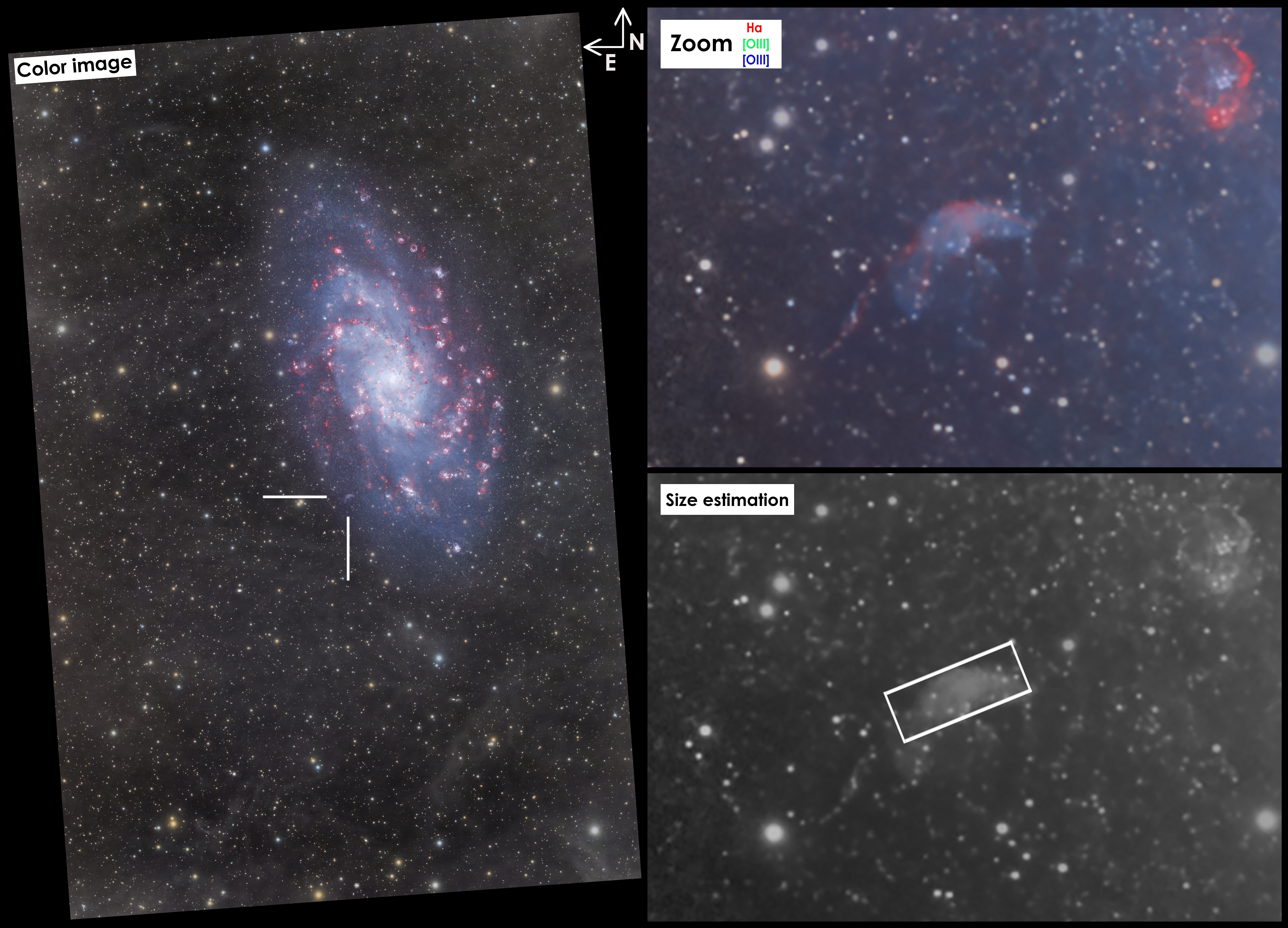}
    \caption{
    {\it Left}: LRGB image of M33.
    B and G represent [OIII], R represents H$\alpha$.
    The outskirts of M33 are mapped according to a classic LRGB composition with no [OIII] or H$\alpha$ added.
    \roig{} is identified by two perpendicular lines.
    {\it Top-right:} Zoom-in of \roig{} from the left panel.
    The HII region \emph{[LHK2017] 318} is at the top right corner.
    {\it Bottom-right}: Same as above in grey scale with a box indicating the size measurement ($0.48\times1.80$\,arcmin).
    }
    \label{figure}
\end{figure*}

The resulting LRGB image is presented in the left panel of Figure~\ref{figure}.
It shows the details in the H$\alpha$ and [OIII] bands of the central part of M33.
We found no emission at these wavelengths on the outskirts of the galaxy.
Foreground Galactic cirrus can be distinguished with LRGB filters across those outer regions.
We refer to the Astrocat Observatory web\footnote{\url{https://astrocat.info/a-deep-triangulum-galaxy/}} to see different versions of the same data.

The top right panel of Figure~\ref{figure} shows a zoom-in of the discovered nebula.
It has an H$\alpha$ surface brightness of $\sim23.9\mast$.
The detection of the same feature using two independent telescopes and several nights rejects the possibility of being an instrumental artefact.

Furthermore, \roig{} is also visible in the deep image of M33 from \cite{M33image_Magrini2000}\footnote{\url{https://www.ing.iac.es/PR/newsletter/news2/m33.pdf}} obtained with the Isaac Newton Telescope.
The nebula is outlined as a bright [OIII] emitter at the top right corner of the second part of their image.
This is a completely independent confirmation of our finding.
The goal of these authors was obtaining a catalogue of planetary nebulae in M33 \citep{M31PNe_Magrini2000}; they did not include our nebula among their candidates and it is not mentioned in the paper.
Indeed, \roig{} is much more extended than the targets they studied.
\cite{outskirtsM33_GaleraRosllo2018} also searched for planetary nebulae at the outskirts of M33, but they do not cover our region of interest.

The centre of the brightest part of this arch-shaped nebula is at $\alpha = 1^{h}34^{m}25^{s}$, $\delta = +30^{\circ}20'17''$ (ICRS).
It has an angular size of $(0.48\times1.80)\pm0.12\,$arcmin (bottom right panel of Figure~\ref{figure}).
Assuming a distance to M33 of $840\pm11\,$kpc \citep{M33dist_Breuval2023}, this translates into a physical size of $(120\times440)\pm30\,$pc.
We detect two fainter filaments that extend even further towards the southeast direction.

\roig{} might be the brightest part of a spherical shell.
In that case, a lower limit of the diameter of this bubble would be $\sim470\pm30$\,pc.
When including the two filaments extending southeast, this shell almost doubles its size.
The angular distance between the nebula and the M33 centre is $21\,$arcmin.
This translates to $\sim5\,$kpc at the aforementioned distance without considering the inclination of the M33 disc.

The southern part of \roig{} coincides with the Wolf-Rayet (WR) star \emph{[NM2011]-J013425.11+301950.3} (\emph{Gaia-DR3-303312480229584384}). 
However, it does not seem to be at the centre of our nebula.
Moreover, \roig{} is much bigger than previous WR nebulae found in M33 \citep[typical diameters of a few tens of parsecs according to][]{M33WRnebula_Drissen1991}.
These two properties are the reasons why we think that this structure might not be the nebula associated with the aforementioned WR.
On the other hand, the high emission of excited oxygen makes us consider \roig{} as a possible supernova remnant.

The real nature of \roig{} is unknown.
We plan to perform new observations and obtain deep spectroscopic data to determine its morphological, chemical and kinematic properties.

\section{Acknowledgements}
We acknowledge Dr. Jorge Garc\'ia-Rojas, Dr. \'Angel L\'opez-S\'anchez, Dr. Ignacio Trujillo, Dr. Kike Herrero and Josep M. Drudis for their helpful advice and comments.
AR thanks Georgina Serven for her help during this project.
RIS acknowledges governments of Spain and Arag\'on, FITE, and Science Ministry (PGC2018-097585-B-C21, PID2021-124918NA-C43).
JA acknowledges the funding by Spanish MCIN/AEI/10.13039/501100011033 and by ESF+ through the PRE2021-100596 grant.

%% the bibliography. The references.bib file was populated from ADS. To
\bibliography{references}{}
\bibliographystyle{aasjournal}

\end{document}